 \documentclass[draft,twocolumn, aps, pre, eqsecnum,amsmath, amssymb,showpacs]{revtex4}

 \usepackage[final]{graphicx}  

 \newcommand{\Nabla}{\mbox{\bf\boldmath $\nabla$}}                              
                                                              
 \renewcommand{\vec}[1]{{\bf #1}}

\begin{document}

 \title{\bf 
 Localized waves whithout the existence of extended waves:\\ 
 oscillatory convection of binary mixtures with strong Soret effect}                                

 \author{D.~Jung and M.~L\"ucke}                                   
 \affiliation{Institut f\"ur Theoretische Physik, Universit\"at des Saarlandes,      
 Postfach 151150, \\ D-66041 Saarbr\"ucken, Germany}                              
   
 \date{\today}                                                                   

 \begin{abstract}
 Spatially confined solutions of traveling convection rolls are determined 
 numerically for binary mixtures like ethanol-water. The appropriate field
 equations are solved in a vertical crossection of the rolls perpendicular 
 to their axes subject to realistic horizontal boundary conditions.
  The localized convective states 
 are stably and robustly sustained by strongly nonlinear mixing and complex
 flow-induced concentration redistribution. We elucidate how
 this enables their existence for strongly negative separation ratios at
 small subcritical
 heating rates below the saddle-node of extended traveling convection rolls 
 where the quiescent fluid is strongly stable.
 \end{abstract}
 
  \pacs{47.20.-k, 47.54.+r, 44.27.+g, 47.20.Ky}
 \maketitle                                                                  
 \vskip2pc


 Many nonlinear dissipative systems that are driven sufficiently far away from 
 thermal equilibrium show selforganization out of an unstructured state: A 
 structured one can appear that is characterized by a spatially extended pattern 
 which retains some of the symmetries of the system \cite{CH93}. Some systems 
 form also spatially confined structures \cite{loc_states}. In particular 
 localized traveling wave (LTW) structures that are embedded at 
 subcritical driving rates in the stable featureless surrounding and that
 compete with the subcritically bifurcated extended traveling wave (TW) pattern 
 have been explored in experiments \cite{exp}, numerical
 simulations of the appropiate field equations \cite{LTW,Y91_NHY96}, and via 
 various model approaches \cite{model,Riecke}.
 The coexistence of the TW state and the quiescent state 
 is a prerequisite for interpreting LTWs as spatial connections of the 
 former to the latter via two fronts. However, this picture coming from
 simple cubic-quintic amplitude equations is too simple for convection
 in binary mixtures: there the strongly nonlinear TWs cannot be described 
 by a power-law expansion in one amplitude \cite{HBL97}. Here we show that 
 LTWs are - independent of TWs - separate nonlinear states.
 
 We predict and describe in quantitative detail stable subcritical LTWs in
 mixtures with strong Soret 
 effect at small heating rates where no extended TW whatsoever -- stable or 
 unstable -- exists \cite{worms}. Furthermore, we elucidate that
 the concentration field and its current are the key to understand how LTW 
 convection is sustained when the quiescent conductive state is strongly 
 stable. The latter is solutally stabilized since the Soret coupling
 between deviations $\delta T$ of temperature and $\delta C$ of concentration
 from their means increases (decreases) the lighter component
 in cold (warm) fluid regions thereby reducing the buoyancy. The most important
 effect is a flow-induced concentration redistribution within
 the LTW: It is the strongly nonlinear mixing by the LTW flow with large Peclet
 numbers that reduces the Soret-generated, stabilizing concentration gradients 
 and that increases {\em locally} the driving buoyancy force 
 to levels which suffice to cause convection in {\em well mixed} fluid with
 strongly reduced concentration variations \cite{discuss}. 


 We have solved the field equations \cite{LBBFHJ98} for convection in binary 
 mixtures like ethanol-water for Lewis number $L$=0.01, Prandtl number 
 $\sigma$=10, and separation ratios \cite{CH93} $-0.25 \leq \psi \leq -0.4$ with 
 a finite-difference method \cite{BLKS95I} in a vertical cross 
 section through the convection rolls perpendicular to their axes. To simulate 
 1D patterns arising in narrow annular channels we applied laterally periodic 
 boundary conditions at $x=0, \Gamma$ with $\Gamma$ up to 160 times the layer 
 height to accomodate the largest LTWs. Spatially extended TW states of 
 different wavelength $\lambda=2\pi/k$ were obtained with $\Gamma=\lambda$. 
 Horizontal boundaries at top and bottom, $z=\pm 1/2$, are realistic, i.e., no 
 slip, perfectly heat 
 conducting, and impermeable. Our control parameter $r=R/R^0_c$ measuring the 
 thermal driving is the Rayleigh number $R$ reduced by the critical one 
 $R^0_c$=1707.762 for onset of convection in a pure fluid.
 To measure the lateral variation of the mixing in oscillatory 
 convective flow we use the mixing number $M(x)$ 
 \begin{equation}
 M^2 = \overline{\langle (\delta C)^2 \rangle} \Big/ \,
 \overline{(\delta C_{cond})^2}\,\, .
 \label{Eq:mixnumber} 
 \end{equation}
 Here overbars imply vertical averaging. Brackets denote temporal averaging, 
 $\langle f \rangle = \langle f(x+v_d t,z,t) \rangle$, 
 over one oscillation period in the frame comoving with the slow drift velocity $v_d$ 
 of the LTW envelope. In a perfectly mixed fluid $M$ vanishes while $M\to 1$ in the 
 conductive state [denoted by the subscript $cond$ in Eq.(\ref{Eq:mixnumber})]
 with its large Soret-induced vertical concentration gradient.

 In Fig.~\ref{Fig:bif1} we show how an increasing Soret coupling strength 
 changes the bifurcation properties of LTW width $l$, maximal vertical flow velocity
 $w_{max}$, frequency $\omega$, and LTW drift velocity $v_d$. The driving interval 
 ($r_{min}, r_{max}$) with stable stationary LTWs moves upwards and 
 its extension increases with $|\psi|$. At its lower end LTWs are for all $\psi$ 
 narrow pulses of universal structure and $l$ of about 5. At the upper 
 end $l$ seems to diverge. LTWs with, say, $l \lesssim$ 7 behave differently than the 
 broad-width LTWs (cf. Fig.~\ref{Fig:struct}) with an
 extended TW like center part between the two limiting fronts. The change in the
 LTW bifurcation diagrams of $\omega$ and $v_d$ versus $r$ in Fig.~\ref{Fig:bif1} 
 reflects this difference. Our simulations indicate that the LTWs of 
 Fig.~\ref{Fig:bif1} are uniquely selected, monostable 
 confined solutions \cite{bistability} and robust against small perturbations.

 With increasing $|\psi|$ TW saddle-node
 location $r_s^{TW}$ and oscillatory threshold $r_{osc}$ move much faster to 
 larger $r$ than the lower LTW band edge $r_{min}$. At $\psi$=-0.08 (not 
 shown here) one has $r_s^{TW} < r_{osc} < r_{min}$ so that there the narrow 
 LTW pulses that have been investigated extensively lie in
 the linearly convectively unstable parameter region (see, e.g., 
 \cite{LBBFHJ98} for discussion and references). 
 However, already at $\psi$=-0.4 almost all LTWs appear {\em below} the TW 
 saddle-node, i.e., ahead of
 the TW nose of Fig.~\ref{Fig:bif2}. Thus, stable LTW convection is driven here
 at heating rates $r$ for which any extended TW convection is impossible.
 Even at $\psi$=-0.25 the relevant LTW parameter combinations of 
 $\omega, r, k_{plateau}$ (big bullets in Fig.~\ref{Fig:bif2}) lie outside 
 and ahead of the TW bifurcation surface. For all $\psi$ investigated here the 
 LTW width seems to diverge with increasing $r$ when its frequency and plateau
 wave number $k_{plateau}$ (big bullets in Fig.~\ref{Fig:bif2}) approaches 
 an $\omega - r - k$ combination of a TW.

 LTWs are sustained by the following complex,
 large-scale concentration redistribution process.
 While traveling from tail to head LTW rolls increase their lateral 
 concentration contrast, i.e., $M$ and with it $v_p$, and $\lambda$ \cite{Mvp}
 (cf. Fig.~\ref{Fig:struct}). 
 Positive "blue" (negative "red") concentration deviation 
 from the global mean is sucked from the top (bottom) boundary layer into
 right (left) turning rolls as soon as they become nonlinear under the trailing
 front. This happens when the vertival velocity $w$ roughly
 exceeds $v_p$ [left arrow in Fig.~\ref{Fig:struct}(b)] so that regions with 
 closed streamlines appear \cite{HBL97,LBBFHJ98}. Within them "blue" ("red")
 concentration is transported predominatly in the upper (lower) part 
 of the layer to the right.
 Mean concentration, on the other hand, migrates mostly to the left 
 along open streamlines that meander between the closed roll regions and that 
 follow the border line between green and yellow in Fig.~\ref{Fig:struct}(a).
 The {\em time averaged} current of $\delta C$ [green lines in 
 Fig.~\ref{Fig:struct} (e)] reflects the mean properties of this  
 transport. Since "blue" and "red" ("yellow") is transported away from
 (towards) the left trailing front mean concentration accumulates there and 
 causes a strong drop of $M(x)$.
 By the same token the leading front's concentration variations and with 
 it $M(x)$ are strongly increased even beyond the conductive 
 state's values. Thus, unlike TWs LTWs do not reach a 
 balance between $\delta C$ injection and 
 advective mixing and diffusive homogeneization on a constant level of small 
 $M$. Rather LTW rolls 
 collaps under the leading front when $v_p$ has grown up to $w$ [right arrow in 
 Fig.~\ref{Fig:struct}(b)]. Thereafter concentration is discharged and sustains 
 a barrier of $\langle \delta C \rangle$ ahead of the leading front. When the 
 mixing
 increases with $w_{max}$ and $r$ so does $l$ since the slower growth of 
 $\delta C$ and $M$ to the higher levels necessary for the leading front's transition
 to conduction requires longer and longer propagation lengths for the rolls.

 The lateral redistribution causes the mean convectively generated $C$-profile
 [green line in Fig.~\ref{Fig:struct}(d)] to extend significantly further into the
 conductive region than the mean convective temperature field (red line). Thus,
 the buoyancy $\langle b \rangle$ 
 [black line in Fig.~\ref{Fig:struct}(d)] is determined in the front regions
 predominantly by the concentration field. This explains ({\it i}) the decrease 
 of buoyancy below conduction levels ahead of the leading front with the 
 associated restabilization of conduction there and ({\it ii}) the increase of 
 $\langle b \rangle$ out of the conductive state under the trailing front and its 
 strong overshoot [cf. Fig.~\ref{Fig:struct}(d)] over the bulk enabling 
 convection growth. 

 That the driving motor for LTW convection is located under the trailing 
 front where $\langle b\rangle, M, v_p, \lambda$ are extremal can also be seen 
 from the relaxation behavior after changing $r$: Not only 
 $\langle b\rangle, M$, and $v_p$ but also 
 $\omega$ and $v_d$ appropriate for the new $r$ are almost instantaneously 
 realized locally under the trailing front while the relaxation of the leading
 front involving also diffusive processes
 takes much longer time, cf., Fig.~\ref{Fig:relax}. Thus there $\omega$ and $v_d$
 is largely selected at the trailing front while the large-scale lateral
 concentration transport and redistribution determines the leading front's
 properties and its location, i.e., the LTW width. 

 To summarize: We have found, analyzed, and explained uniquely selected stable 
 subcritical LTWs in mixtures with strong Soret effect
 at small heating rates where extended TWs are not possible, i.e., below the
 saddle node bifurcation of the latter. These strongly nonlinear LTWs are 
 self-consistently sustained by a concentration redistribution such that  
 flow-induced mixing increases locally the buoyancy to levels that
 suffice to drive well mixed  fluid flow.



 \begin{figure*}
 \caption{Bifurcation properties of LTWs (symbols) and TWs (lines) for different
 separation ratios $\psi$: (a) Full width $l$ of LTWs at half maximum of the 
 envelope of the vertical velocity field $w$ [blue line in 
 Fig.~\ref{Fig:struct}(b)].
 For small filled symbols $l$ kept on growing slowly. (b) Maximal vertical flow 
 velocity $w_{max}$. (c) Frequency; for LTWs in
 the frame comoving with $v_d$.
 (d) Drift velocity $v_d$ of LTWs. Phase velocity is always to the
 right, positive, and much larger. Thick lines in (b, c) denote 
 TWs with saddle-node 
 wave number $k_s^{TW} \simeq \pi$. Thin lines are TWs with plateau wave 
 number $k_{plateau}$ of the last LTW before the $l\to\infty$ transition 
 from LTW to TW. The frequency of the latter is determined in the frame comoving 
 with the last LTW. Unstable TWs (dashed lines; determined with a control
 method) bifurcate subcritically with large Hopf frequency $\omega_H(k)$ at 
 $r_{osc}(k)$ out of the conductive state and become stable (solid lines) at
 the saddle-node $r_s^{TW}(k)$ when lateral periodicity is imposed with
 $\Gamma=\lambda=2\pi/k$. \label{Fig:bif1}}
 \end{figure*}
 \begin{figure*}
 \caption{Broad LTW of $l$=17.4: (a) Snapshot of 
 concentration deviation $\delta C$ from global mean (light green/yellow) in
 a vertical cross section of the layer. (b) Snapshots of lateral wave
 profiles at midheight, $z$=0, of $\delta C$ (green), vertical 
 velocity $w$ (blue), and its envelope. At the arrows $w_{max}=v_p$.
 (b) Mixing number $M$ (green), Eq. (\ref{Eq:mixnumber}) and phase velocity 
 $v_p$ (black) of
 nodes of $w(z=0)$ in frame comoving with $v_d$. The variation of 
 $\lambda(x)=2\pi\;v_p(x)/\omega$  is the same
 since the LTW frequency $\omega$ is a {\em global} constant. Thin 
 (thick) bullet marks smallest (plateau) wavelength for discussion. 
 (d) Time averaged deviations from the conductive state at $z$=-0.25 for 
 concentration (green), temperature (red), and their sum ($\langle b \rangle$)
 measuring the convective contribution to the buoyancy. 
 (e) Streamlines of time averaged concentration current 
 $\langle \vec{J}\rangle = 
 \langle \vec{u}\delta C - L\Nabla (\delta C -\psi \delta T)\rangle$ 
 (green) and velocity field $\langle \vec{u} \rangle$ (blue). The latter results
 from $\langle b \rangle$ and documents roll shaped contributions of 
 $\langle \vec{u} \rangle \langle \delta C \rangle$ to $\langle \vec{J}\rangle$
 under the fronts and the associated $\langle \delta C\rangle$ redistribution. 
 Thick blue and green arrows indicate $\langle \vec{u} \rangle$ and
 transport of positive $\delta C$ (alcohol surplus), respectively. Thus,
 in the lower half of the layer negative $\delta C$ (water surplus) is tranported
 to the right. Parameters are $\psi$=-0.35, $r$=1.346. \label{Fig:struct}}
 \end{figure*}
 \begin{figure*}
 \caption{TW and LTW oscillation frequencies over the $k-r$ plane. 
 Extended TWs (nose shaped surface) 
 start as unstable solutions with $\omega_H(k)$ at upper thick line -- its
 projection onto the $k-r$ plane is the threshold curve $r_{osc}(k)$. Dent in
 upper nose part is related to appearence of closed
 $w$-streamlines when $w_{max} \simeq v_p$ causing $\delta C$-anharmonicities and 
 the breakdown of small-amplitude expansions \cite{HBL97,LBBFHJ98}.
 With decreasing $\omega$ TWs become stable in periodic systems of 
 $\Gamma=2\pi/k$ at the saddle-node line
 (dotted outermost perimeter of the nose) located at $r_s^{TW}(k)$. For large
 $\Gamma$ TWs become stable at the slightly larger Eckhaus boundary $r_E^{TW}(k)$ 
 \cite{Eckhausinst}. TWs with 
 $k_s^{TW} \simeq \pi$ of the tip of the nose (plateau wave number 
 $k_{plateau}$ of the last
 LTW before its $l\to\infty$ transition) are marked - as in Fig.~\ref{Fig:bif1} - 
 by thick (thin) curves on the nose surface. Thin dotted 
 horizontal lines denote LTWs: they contain for fixed $\omega$ a broad band of 
 wave numbers extending from the largest ones, $k_{max} \simeq$ 4.3 
 (small bullets as in Fig.~\ref{Fig:struct}), to small ones outside the plot 
 range. Plateau wave numbers $k_{plateau}$ are marked as in 
 Fig.~\ref{Fig:struct} by big bullets that appear only where $l$ is 
 sufficiently large. \label{Fig:bif2}}
 \end{figure*}
 \begin{figure*}
 \caption{Thick full (dashed) lines show front relaxation over a limited time
 interval after changing the
 driving from $r$=1.41 to 1.43 (1.38). Thin gray (black) lines are world lines 
 of  $w$-nodes (front positions) of the original LTW 
 ($r$=1.41, $\psi$=-0.4, $l$= 20). \label{Fig:relax}}
 \end{figure*}


\begin{thebibliography}{999}
\bibitem{worms}
Recently 2D subcritical localized states (so-called "worms") that could be
generated from thermally driven fluctuations perturbing the conductive state
have been seen in an 
electroconvecting nematic liquid crystal while experiment and theory indicate 
that extended convection bifurcates supercritically \cite{expworms}. A weakly 
nonlinear model showed also subcritical worm-like solutions in the presence of
a conjectured additional field that is advected obliquely from both sides 
towards the worms \cite{RG99RR01}.
\bibitem{bistability}
 The larger LTWs of the bistable ones found in relative short systems 
 ($\Gamma =$ 20 and 40) \cite{LTW} are presumably stabilized by head-tail
 interactions.
\bibitem{Mvp}
 $M$ and $v_p= \omega \lambda /2 \pi$ of stationary \cite{HBL97} and transient
 \cite{FL} TW solutions
 are practically proportional to each other. This roughly holds also locally in
 LTWs where $M, v_p, \lambda$ vary spatially while $\omega$ is constant in the
 comoving frame.
\end{thebibliography}
\end{document}